\documentclass[twocolumn, epjc3]{svjour3}
\smartqed  

\usepackage{amsmath, amssymb, mathrsfs}
\usepackage{mathtools}
\usepackage{slashed}
\usepackage[english]{babel}
\usepackage{graphicx,color,psfrag}
\usepackage{tikz}
\usepackage{pgfplots}
\usepackage{placeins}
\usepgfplotslibrary{fillbetween}
\pgfplotsset{compat=1.14}
\usepackage{dirtytalk}
\usetikzlibrary{spy,backgrounds}

\input{couleurs.sty}

\newcommand{\dd}{\ensuremath{\mathrm{d}}}

\newcommand{\ket}[1]{\ensuremath{\left| #1 \right>}}
\newcommand{\bra}[1]{\ensuremath{\left< #1 \right|}}
\newcommand{\bracket}[2]{\ensuremath{\left< #1 \right| #2 \left . \right>}}
\newcommand{\Tr}[1]{\ensuremath{\mathrm{Tr}\left ( #1 \right )}}

\usepgfplotslibrary{external}
\tikzset{external/system call={lualatex --shell-escape
		\tikzexternalcheckshellescape --halt-on-error --interaction=batchmode
		--jobname "\image" "\texsource"}}
\usetikzlibrary{shapes.misc}
\usetikzlibrary{plotmarks}

\tikzset{cross/.style={cross out, draw=black, minimum size=2*(#1-\pgflinewidth), inner sep=0pt, outer sep=0pt},
	cross/.default={2.2pt}}
\usetikzlibrary{pgfplots.groupplots}

\newcommand{\modif}[1]{{\color{black}#1}}

\journalname{Eur. Phys. J. C}

\begin{document}
\title{\Large Open-Boundary Conditions in the Deconfined Phase}


\author{Adrien Florio\thanksref{e1,addr1}
        \and
        Olaf Kaczmarek\thanksref{e2,addr2,addr3} 
        \and
        Lukas Mazur\thanksref{e3,addr3}
}

\thankstext{e1}{e-mail: adrien.florio@epfl.ch}
\thankstext{e2}{e-mail: okacz@physik.uni-bielefeld.de}
\thankstext{e3}{e-mail: lmazur@physik.uni-bielefeld.de}

\institute{Institute of Physics, Laboratory of Particle Physics and Cosmology, Ecole Polytechnique F\'ed\'erale de Lausanne,  CH-1015 Lausanne, Switzerland\label{addr1}
          \and
            Key Laboratory of Quark \& Lepton Physics (MOE) and Institute of Particle Physics, Central China Normal University, Wuhan 430079, China\label{addr2}
          \and
          Fakult\"at f\"ur Physik, Universit\"at Bielefeld,  33615 Bielefeld, Germany\label{addr3}
}


\date{Received: date / Accepted: date}

\maketitle

\abstract{In this work, we consider open-boundary conditions at high temperatures, as they can potentially be of help to measure the topological susceptibility. In particular, we measure the extent of the boundary effects at $T=1.5T_c$ and $T=2.7T_c$. In the first case, it is larger than at $T=0$ while we find it to be smaller in the second case. The length of this \say{boundary zone} is controlled by the screening masses. We use this fact to measure the scalar and pseudo-scalar screening masses at these two temperatures. We observe a mass gap at $T=1.5T_c$ but not at $T=2.7T_c$. Finally, we use our pseudo-scalar channel analysis to estimate the topological susceptibility. The results at $T=1.5T_c$ are in good agreement with the literature. At $T=2.7T_c$, they appear to suffer from topological freezing, which prevents us from providing a precise determination of the topological susceptibility.
}

\section{Introduction}

In general, finite-size systems differ from their infinite-volume counterpart. One of the most simple examples is the  "particle-in-a-box" whose momenta are quantised. Not only the compactness, but also the boundary conditions affect the system. There, different choices lead to different quantisation conditions. The only restriction on such choices is that the infinite volume physics needs to be recovered in the thermodynamic limit. This requirement satisfied, the only remaining differences are related to the convergence to the infinite volume limit. When the system is discretised, discretisation effects may also vary between different types of boundary conditions.

In some circumstances, such differences may be used as algorithmic tools to improve numerical simulations. A typical example of this is the use of open-boundary conditions (OBC) in lattice QCD, which have been introduced in \cite{Luscher2011} as  means to reduce autocorrelations of the topological charge. These autocorrelations  become critical as the continuum is approached. and are signaled by the freezing of gauge field ensembles in given topological sectors.
In this example, instead of considering QCD with periodic boundary conditions (PBC), which leads to a discrete topological charge

\begin{equation}
Q=\int \dd^4 x\,  q(x)=\frac{1}{32 \pi^2}\int \dd^4 x\, \epsilon^{\mu\nu\rho\sigma}\Tr{G_{\mu\nu}G_{\rho\sigma}},
\label{eq:topCharge}
\end{equation}
the idea is to impose OBC in at least one of the directions. In this system, $Q$ spans a continuum range of value. This then lifts the topological barrier responsible for the topological freezing and improves the sampling of the configuration space.

Having small autocorrelations is crucial to keep control of the statistical errors in Monte Carlo simulations \cite{Wolff:2003sm,Schaefer2011}. A poor sampling of
the topological charge affects in principle all observables, leading to finite volume effects (see  \cite{Aoki:2007ka,Bietenholz:2016ymo} for practical examples).
The situation is partially improved when considering QCD in the deconfined phase. For $T>T_c$, the order parameter which quantifies the variance of the topological charge, i.e. the topological susceptibility $\chi=\frac{\langle Q^2\rangle}{V}$, decreases with $T$. At asymptotically-high temperatures, it is even suppressed as $T^{-7}$ \cite{Gross:1980br}. Nonetheless, for moderate temperatures, $Q\neq 0$ configurations still contribute in a non-negligible way to the path-integral.
In this context, OBC  may also be of interest at non-zero temperatures\footnote{For a very exploratory study, see \cite{Burnier:2017osu}.}. However, before being able to use them systematically, an analysis of the influence of temperature  on the boundary effects remains to be done. This is the content of this study, which focuses on pure $SU(3)$ gauge theory, as dynamical matter is not expected to drastically change the results.

In section \ref{sec:OBC}, we recall known facts about OBC and discuss our datasets and methodology. Then,  in the spirit of the zero temperature analysis of \cite{Husung:2017qjz}, we investigate in section \ref{sec:bczone} the typical length over which the boundary effects propagate, the \say{boundary zone}. We observe a noticeable temperature dependence. These differences can be understood in terms of the temperature dependence of the lightest propagating states' screening masses, which we study in section \ref{sec:screening_mass}.  As a by-product, we report in section \ref{sec:top_susc}
an extraction of the topological susceptibility from our rather large volumes simulations. We finally discuss our results in section \ref{sec:conclu}.

\section{Open-Boundary Conditions and Setup}
\label{sec:OBC}
Conventional lattice QCD simulations use (anti-)periodic boundary conditions in all directions, for the obvious reason that they minimise boundary effects. In this study, we consider the use of OBC in one of the spatial directions (taken for definiteness to be the $x$ direction). This amounts setting the field-strength tensor to zero outside the lattice. In this case the Wilson action reads \cite{Luscher2011}

\begin{equation}
  S^{OBC}=-\frac{\beta}{3}\sum_P w(P)\Tr{\mathbf{1}-P},
  \label{eq:acSOBC}
\end{equation}
where the sum runs over all the plaquettes

\begin{equation}
  P_{\mu\nu} = U_{\mu}(n)U_\nu(n+\hat{\mu})U_{\mu}(n+\hat{\nu})^\dagger U_\nu(n)^\dagger
\end{equation}
 whose corners are in the interval $[x=0,x=N_x-1]$. The $U$'s are the usual link variables and the quantity $w(P)$ is an integration weight
\begin{equation}
w(P)=\begin{cases}1\text{ if } P\in \text{bulk} \\ \frac{1}{2} \text{ if } P\in x\text{-face}\end{cases} .
\label{eq:SOBC}
\end{equation}
A bulk point is a point in the interval $[1,N_x-2]$. A plaquette is on a $x$-face if it is not oriented along $x$ and all of its corners are at $x=0$ or $x=N_x-1$. As shown in \cite{Luscher2011}, the continuum limit of this theory has a trivial topology in field space; all the admissible fields are connected by local gauge transformations.

Such boundary conditions break translational invariance and introduce boundary effects. These effects may be understood as the propagation of excited states from the boundary. Here we summarise the core of the argument, following \cite{Guagnelli:1999zf,Bruno2016,GarciaVera2017}.

For the sake of clarity, let us first recall the argument for OBC in the time direction; it straightforwardly transposes to OBC in the $x$ direction. To quantise our Euclidean theory, we write down a transfer matrix $\hat{T}=e^{-\hat{H}}$ with $\hat{H}$ the Hamiltonian,  the Euclidean equivalent of the evolution operator. It  evolves states between temporal slices. In particular, going from the state $\ket{\gamma_i}$ at $t=0$ to the state $\ket{\gamma_f}$ at $t=T$, and given an operator $O$ inserted at $t$, we can write

\begin{equation}
  \langle O\rangle_{OBC}=\frac{1}{\mathcal{Z}}\bra{\gamma^f}\hat{T}^{-(T-t)}O(t)\hat{T}^{-t}\ket{\gamma^i} .
\end{equation}
To label our basis of states, we use the lattice version of the translation operators and get a basis consisting of $\ket{E_n(\vec{p})}$, with $n$ labelling extra quantum numbers and $\vec{p}$ the momentum eigenstates. Inserting a complete basis of states, we can then write

\begin{align}
  \langle O\rangle_{OBC}&=\frac{1}{\mathcal{Z}}\sum_{n,\vec{p},m,\vec{q}}\gamma^i_{n,\vec{p}} \gamma^{f*}_{m,\vec{q}}\notag\\
  &\cdot\bra{E_m(\vec{q})}\hat{T}^{-(T-t)}O(t)\hat{T}^{-t}\ket{E_n(\vec{p})}\\
  &=\frac{1}{\mathcal{Z}}\sum_{n,\vec{p},m,\vec{q}}\gamma^i_{n,\vec{p}} \gamma^{f*}_{m,\vec{q}}
  e^{-(T-t)E_m(\vec{q})}e^{-tE_n(\vec{p})}\notag\\
  &\cdot\bra{E_m(\vec{q})}O(t)\ket{E_n(\vec{p})},
\end{align}
 with $\gamma^{i,f}_{n,\vec{p}}=\bracket{E_n(\vec{p})}{\gamma^{i,f}}$. Now we see that the main contribution comes from  the state with smallest energy. We then have a tower of exponentially suppressed corrections. More explicitly, using the fact that the main contribution to $\mathcal{Z}$ is $\gamma^i_{0} \gamma^{f*}_{0}e^{-E_0T}$ (obtained by setting $O(t)=\mathbf{1}$ in our expansion), we find

\begin{align}
\langle O\rangle_{OBC}=\bra{0}O\ket{0}&+\alpha_1 e^{-(E_1-E_0)t}\\
&+\beta_1 e^{-(E_1-E_0)(T-t)}+\dots\notag .
\end{align}
with $\alpha_1$ and $\beta_1$ some matrix elements.

In other words, OBC do not project out directly on the vacuum state but are affected by states which propagate from the boundary. We also see that, at least in some limits, the corrections should be dominated by an exponential decay in the lightest state. We will take advantage of this in section \ref{sec:screening_mass}. Note  that this argument can be generalised to two-point functions \cite{Bruno2016} and higher-point functions.

In the case of OBC in the $x$ direction, the previous analysis can be repeated by replacing the slicing in the $t$ direction by a slicing in the $x$ direction when quantising the system. Then $H$ and $\hat{P}_x$ exchange roles, with $\hat{P}_x$ the translation operator in $x$. Modulo this, the derivation goes through.

\begin{table}
\centering
\begin{tabular}{|c|c|c|c|c|c|c|c|}
\hline
 $N_s^3\times N_t$ & $\beta$ &$a \mathrm{[fm]}$& $l\over \sqrt{8t_0}$&$\frac{T}{T_c}$ &$n_{OBC}$ &$n_{PBC}$\\
\hline
\hline
 $64^3\times 6$ & $6.139$&$0.074$ &$10.0$& $1.5$ &$257$ & $442$\\
\hline
 $88^3\times 8$ & $6.335$&$ 0.056$&$10.4$ & $1.5$ &$430$ & $533$\\
\hline
 $112^3\times 10$ & $6.498$&$ 0.045$&$10.7$ & $1.5$ &$246$ & $155$\\
\hline
 \multicolumn{4}{|c|}{Continuum extrapolations for OBC} & $1.5$ &\multicolumn{2}{c|}{-}\\
\hline
 \multicolumn{4}{|c|}{Continuum extrapolations for PBC} & $1.5$ &\multicolumn{2}{c|}{-}\\
\hline
 $64^3\times 4$ & $6.253$ &$0.063$&$8.52$ &$2.7$&$533$&$512$ \\
\hline
 $88^3\times 6$ & $6.55$ &$0.042$&$7.81$ &$2.7$&$318$&$532$ \\
\hline
 $120^3\times 8$ & $6.778$ &$0.031$&$7.86$ &$2.7$&$533$&$532$ \\
\hline
 \multicolumn{4}{|c|}{Continuum extrapolations for OBC} & $2.7$ &\multicolumn{2}{c|}{-}\\
\hline
 \multicolumn{4}{|c|}{Continuum extrapolations for PBC} & $2.7$ &\multicolumn{2}{c|}{-}\\
\hline
 $102^3\times6$ & $6.64$ & $0.037$&$7.98$ &$3.0$ &$479$ &$472$ \\
\hline
\end{tabular}
\caption{Lattices used in this study. The spatial size of the lattice is denoted $l$. }
\label{tab:conf}
\end{table}

To measure the topological correlators, we used the gluonic definition of the topological charge density.
It requires some smoothing of the gauge fields, which was performed by using the gradient flow \cite{Luscher:2010iy}. The fundamental gauge fields $A_\mu(x)$ are evolved to finite flow-time $\tau$, $B(x,\tau)$, using the flow equation
\begin{equation}
    \dot{B}_{\mu}(x,\tau)=D_{\nu}G_{\nu\mu}(x,\tau),\,\,\,\,\,\,B_{\mu}(x,\tau)|_{\tau=0}=A_{\mu}(x),\label{eq:gradientFlow}
\end{equation}
\begin{equation}
      D_{\mu}=\partial_{\mu}+\left[B_{\mu}(x,\tau),\,\,\mathord{\cdot}\,\,\right].
\end{equation}
The associated smearing radius is  $\sqrt{8\tau}$. It is implemented on the lattice by using the standard Wilson gauge action (Wilson flow).
The integration is done using a third order Runge-Kutta algorithm with a step-size of $0.01$, which was tested to be small enough for the lattice parameters of this study.

The configurations we used are listed in table \ref{tab:conf}.
The quenched configurations were generated using a heat bath and an overrelaxation algorithm. One update consists of one heat bath
and four overrelaxation steps. To make sure that the configurations are sufficiently thermalised we discard configurations from the first $4000$ iterations.
Configurations are measured  every $500$ Monte Carlo steps to minimise the autocorrelations.
Working with flowed configurations, we use the scale $t_0$ with the interpolation given in \cite{Francis:2015lha} to convert to physical units.
The statistical  errors were estimated by using Jackknife resampling.

To compute the topological charge and energy density we used the clover-shaped field strength tensor
\begin{equation}
    G_{\mu\nu}(x)=\left(\frac{-i}{8a^{2}}\left(Q_{\mu\nu}(x)-Q_{\nu\mu}(x)\right)\label{eq:clover_repr}\right)_{AH},
\end{equation}
where $AH$ is the projection on the traceless antihermitian part and $Q_{\mu\nu}$ is defined as
\begin{align}\label{eq:Q_mu_nu}
Q_{\mu\nu}(x)=U_{\mu,\nu}(x)+&U_{\nu,-\mu}(x)\\
&+U_{-\mu,-\nu}(x)+U_{-\nu,\mu}(x),\notag
\end{align}
with the plaquette discretisation $U_{\mu,\nu}$.

\section{Boundary Zone}
\label{sec:bczone}
As explained in section \ref{sec:OBC}, the presence of a boundary affects observables in the bulk, at least close to the boundary. The length of this \say{boundary zone} depends on how the observables couple to the lightest propagating states. To  quantify this effect and in order to compare it to the
zero temperature case, we adopted the method of \cite{Husung:2017qjz}. We compute the value of the clover action density as a function of the distance to the boundary and extract the length of the boundary zone, i.e. the length over which this observable is significantly different from its bulk value.
In more detail, for lattices with OBC in the $x$ direction and some operator $O$, we define its sub-average at a distance $r$, inside a sub-volume of size $(N_x-2r)\times N_y\times N_z\times N_t$ from the boundary, by

\begin{align}
O^r=\frac{1}{N_yN_zN_t}&\frac{1}{ \left (N_x-2r\right )}\label{eq:av_r}\\
&\sum_{y=0}^{N_y-1}\sum_{z=0}^{N_z-1}\sum_{t=0}^{N_t-1} \sum_{x=r}^{N_x-r-1} O(x,y,z,t) ,\notag
\end{align}
with $0\leq r < N_{x}/2-1$. For $r=0$, we expect the strongest dependence on the boundary excitations. By studying the $r$-dependence, we can then characterise the typical size of the boundary contamination.

At non-zero temperature, the clover action density leads to two independent gluon condensates \cite{Datta:2015bzm,Wandelt:2016oym}
\begin{equation}
   E_{st}=\frac{1}{4} G^a_{0i}G^a_{0i}, \qquad  E_{ss}=\frac{1}{4} G^a_{ij}G^a_{ij} ;
\end{equation}
a \say{magnetic condensate}  $E_{ss}$ and an \say{electric condensate} $E_{st}$.

\begin{figure*}
\centering
  \includegraphics{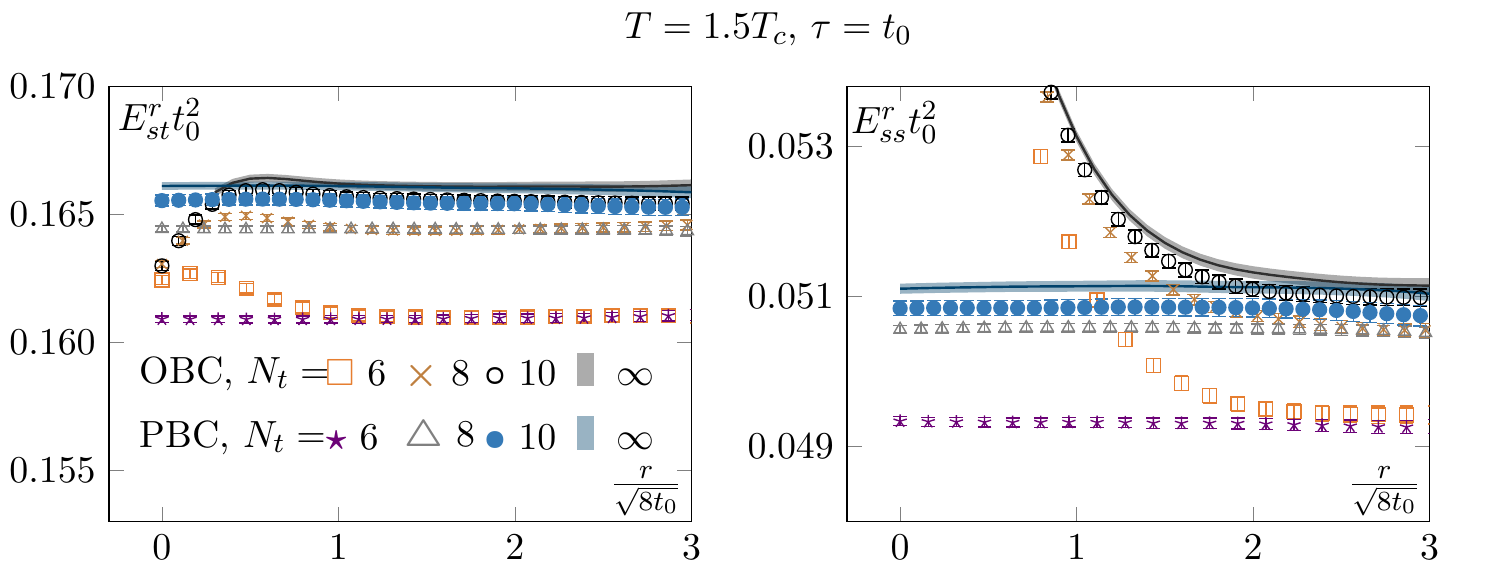}
  \caption{Clover action density as a function of the distance from the open-boundary, at $T=1.5T_c$ configurations and reference flow-time $t_0$. \textbf{Left:} Electric component. \textbf{Right:} Magnetic component. Both components show a consistent scaling to the continuum and an agreement between OBC and PBC in the bulk. Also, we see the effect of the open-boundary. The fact that the two components are not equal is a finite temperature effect (see y-axis). We also see that they do not couple to the same boundary states.}
  \label{fig:clover}
\end{figure*}

In figure \ref{fig:clover}, we show both densities at the reference flow-time $t_0$ for our different configurations at $T=1.5T_c$. All temperatures used in our study behave in a qualitatively similar way. First, we see as expected the existence of a boundary zone and an agreement between OBC and PBC in the bulk of the lattices, i.e. when $r$ is sufficiently large to suppress the effects of the boundary on the sub-volume. Then we see that the component which displays the largest boundary zone is $E_{ss}$. The reason is that it couples to a lighter state than $E_{st}$.

\begin{figure}
\centering
  \includegraphics{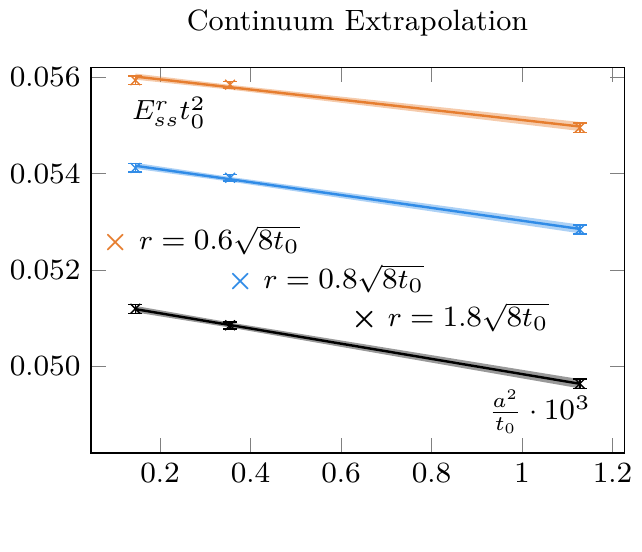}
    \caption{Continuum extrapolation of the magnetic clover density for three different radii $r=0.6,0.8,1.8\sqrt{8t_0}$. }
  \label{fig:lim_corrs}
\end{figure}

To  compare different results in all fairness, we proceed to a continuum extrapolation of both condensates. In figure \ref{fig:lim_corrs}, we show this continuum extrapolation for $T=1.5T_c$ and three different radii. As reported in \cite{Luscher:2010iy}, the region close to the boundary is affected by linear lattice spacing artifacts  when Wilson's action is used without further improvements. We evade this complication by computing our continuum extrapolation only in the region where the $O(a)$ corrections are negligible. This region turned out to be large enough for all purposes of this study.

\begin{figure*}
\centering
  \includegraphics{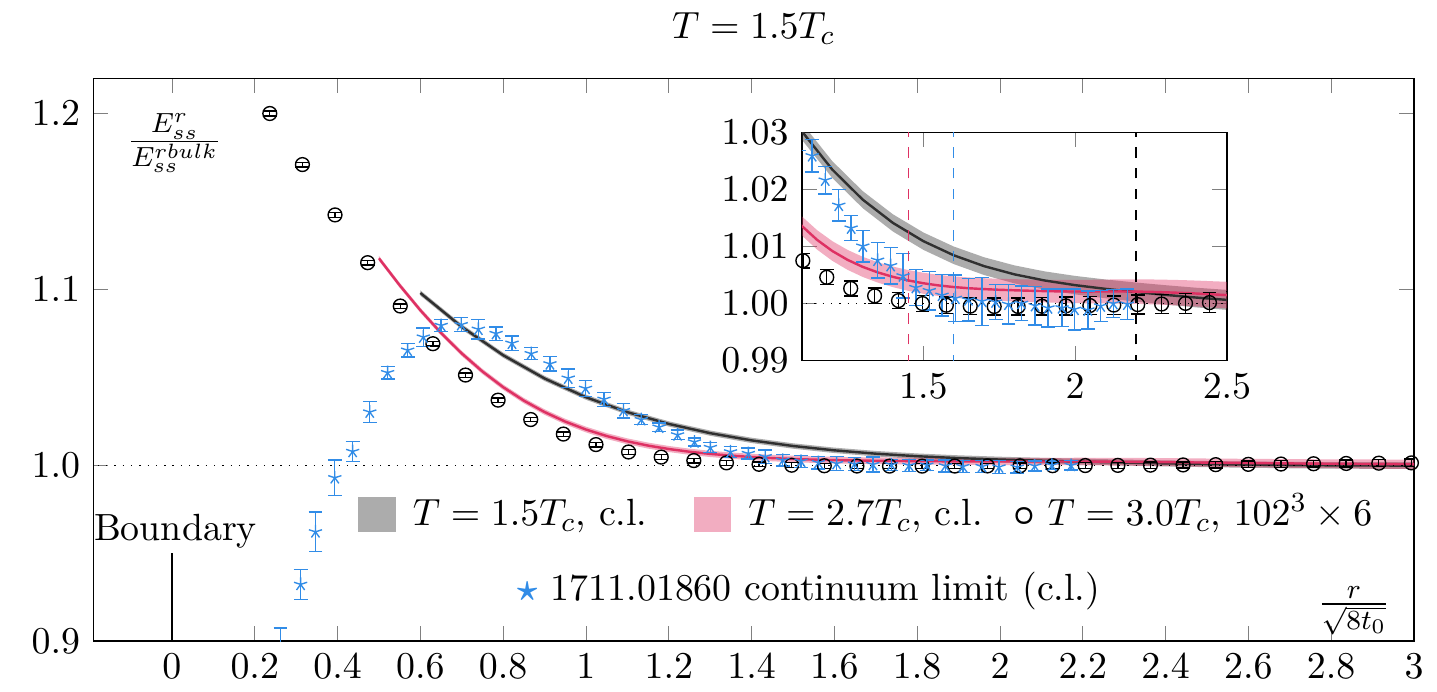}
  \caption{Comparison of the normalised clover action between different temperatures. We report in this figure the zero temperature results of \cite{Husung:2017qjz} in blue. We observe the length of the boundary zone to depend on temperature. At $1.5 T_c$, we see that the boundary effects propagate over a larger distance than at zero-temperature. We take as a conservative estimate of the boundary zone at $1.5T_c$ a length of \modif{$l_b^{1.5T_c}\approx2.2\sqrt{8t_0}$} (black dashed line).
  This has to be compared with the $l_b^{0}\approx 1.6\sqrt{8t_0}$ of \cite{Husung:2017qjz} (blue dashed line). For higher temperatures, the boundary zone gets smaller again. At $2.7T_c$ we estimate it to be of length  \modif{$l_b^{2.7T_c}\approx1.45\sqrt{8t_0}$}.}
  \label{fig:bczone}
\end{figure*}

Different temperatures are compared in figure \ref{fig:bczone}, together with the zero temperature result of \cite{Husung:2017qjz}. In this plot, we show the energy density normalised to its bulk value. We see that the length of the boundary zone depends on temperature. At $1.5 T_c$ we find it to be about $50\%$ larger than at zero temperature while we find it reduced by $20\%$ at $2.7 T_c$, consistently with our fixed lattice spacing results at $3 T_c$. This is also consistent with the temperature dependence of the screening masses. Actually, the behaviour of the observables in the boundary zone gives a handle on these screening masses, which will be discussed in the next section.

\section{Screening Masses}
\label{sec:screening_mass}

As explained in section \ref{sec:bczone}, the boundary effects are controlled by the masses of the propagating states in the theory. In pure $SU(3)$ gauge theory at finite temperature, these are the screening masses \cite{Nadkarni:1986cz}.

In this section, we will take advantage of the boundary effects to extract the lightest scalar and pseudo-scalar screening masses. In particular, as the lightest scalar mass is expected to be the lightest state in our system, its value controls the length of the boundary zone of section \ref{sec:bczone}.

\subsection{Scalar Screening Mass}
\label{sec:scalar_mass}

The strong boundary contamination seen in the $E_{ss}$ channel in figure \ref{fig:clover} suggests that it might be an appropriate probe to extract the scalar screening mass $m_{0^+}$, which will correspond to the lowest screening mass of the state which couples to $E_{ss}$. At zero temperature, it would be the lowest glueball state. Such a strategy was used in \cite{Chowdhury:2014kfa,Chowdhury:2014mra} to extract glueball masses.

To make $E_{ss}$ ultraviolet (UV) finite and be able to take the continuum limit, we study it at some finite flow-time. To have  good control of our errors, we perform a simultaneous fit of the type~
\begin{equation}
  E_{ss}(r)=\alpha \exp\left (-m_{0^+} r\right) +\beta + \gamma a^2
\end{equation}
with $r$ a radius in the boundary zone (see section \ref{sec:bczone}). The constant $\beta$ has to reproduce the continuum bulk value and the $\gamma$ factor encodes the $a^2$ finite lattice spacing corrections.

\begin{figure*}
\centering
  \includegraphics{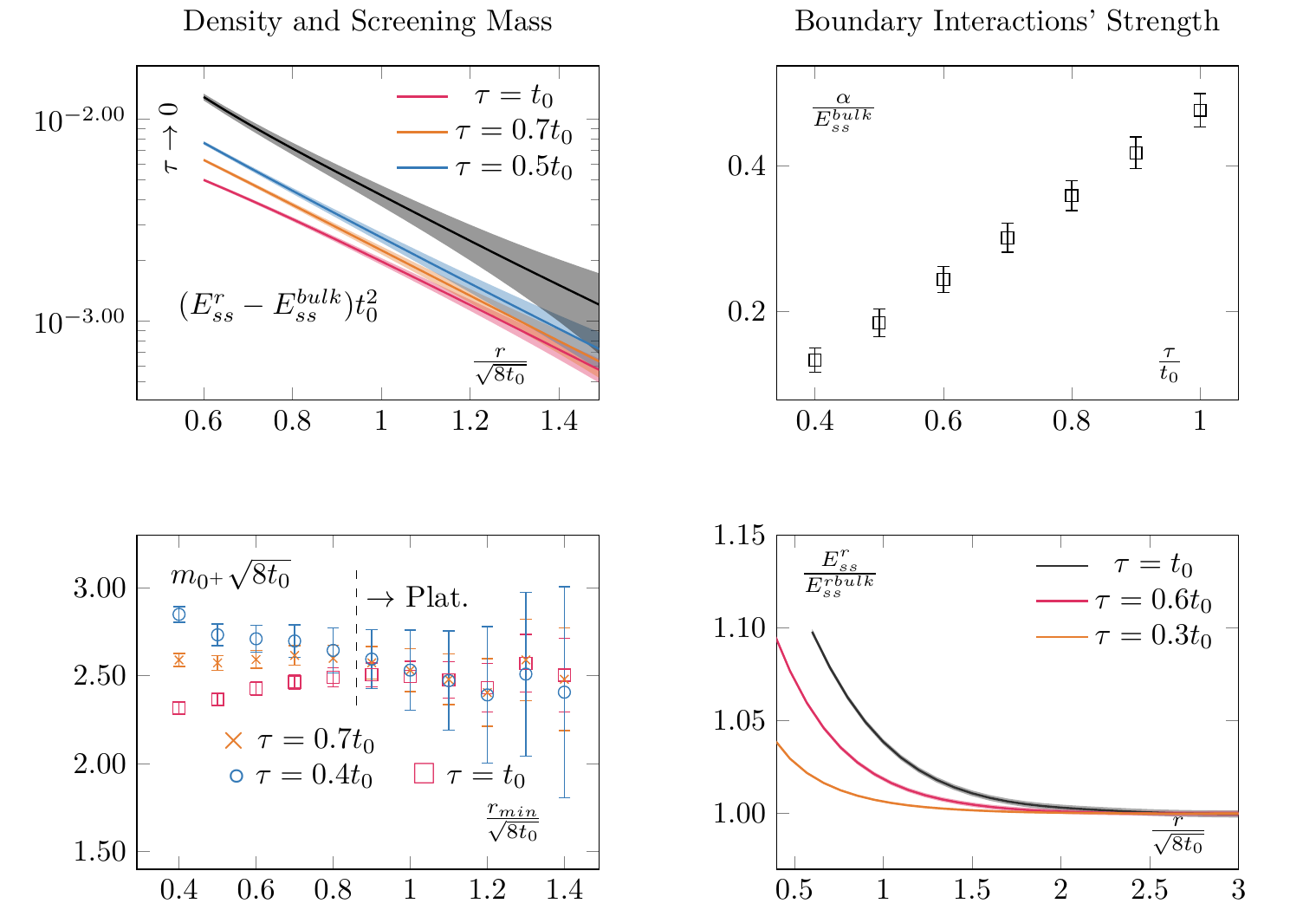}
  \caption{\textbf{Top left:} Continuum extrapolated $(E^{r}_{ss}-E^{bulk}_{ss})$ for different flow-times together with its $\tau\to 0$ limit. Already qualitatively, one can see that there is an exponential decay, whose exponent  does not seem to be sensitive to the flow-time, whilst its prefactor does. \textbf{Bottom left:} Extraction of the effective mass as a function  of the minimal radius used in the exponential fit. We see that when the parameter saturates to a plateau, different flow-times lead to the same prediction, as expected. Note that our errors seem to be overestimated for large $r_{min}$; we do not correct for this. \textbf{Top right:} Normalised prefactor of the exponential. This quantifies the interactions with the boundary states and increases with flow-time. This is due to the smoothing effect of the flow evolution; generically it increases the overlap between states. \textbf{Bottom right:} Corresponding effect on the boundary zone, its length increases with the flow-time as the bulk states interact more and more strongly with the boundary states.}
  \label{fig:screening_proc}
\end{figure*}

We look for an intermediate range $r\in[r_{min},r_{max}]$ of values where we can extract a candidate mass $m_{0+}$. On the left-hand side of figure \ref{fig:screening_proc} we show the behaviour of $\Delta E_{ss}^r t_0^2$ for different flow-times (top panel) together with the extraction of the screening mass for different $r_{min}$ and different flow-times. We also checked that the results were not sensitive to the choice of $r_{max}$.

The extracted screening mass should be flow-time independent, being the mass of some states (the flow evolution will mix different operators but not change the operator basis), and we see that within our precision it is. Outside the plateau region, the masses differ but they do match once a plateau is reached.
Typically, small flow-times lead to a worse signal. The reason lies in the smoothing effect of the flow. For larger flow-times, the errors are reduced and generally speaking overlap between states increases, as do their matrix elements. We can verify this by looking at the prefactor $\alpha$ of our exponential fit, normalised by the bulk value. This quantifies the strength of the interaction with the $0^+$ boundary state. We extract it using the same procedure as for the screening mass and report its flow-time dependence on the top panel on the right-hand side of figure \ref{fig:screening_proc}. As expected, we see it growing with the flow-time. This also explains the behaviour of the boundary region as a function of the flow-time, which is shown in the lower panel on the right-hand side of figure \ref{fig:screening_proc}. The more we flow, the stronger the interaction with the boundary gets and the larger the boundary zone becomes. This suggests that upon a good knowledge of the flow dependence of the observable under consideration, smaller flow-times are advantageous with respect to the boundary contaminations.

In this spirit, it is also instructive to perform the same mass extraction in the limit of zero Wilson flow. It serves two purposes. First, it allows  checking the robustness of our results. Then, since $E_{ss}$ is directly related to the energy-momentum tensor $T_{\mu\nu}$, taking the zero flow-time limit provides a properly renormalised observable. This  would, for example,  be required to extract any running quantities, such as the matrix elements encoded in  $\alpha$.
More precisely let us consider \cite{Suzuki:2013gza}
\begin{align}
  U_{\mu\nu}(x,\tau)&=G_{\mu\rho}^a(x, \tau)G_{\nu\rho}^a(x, \tau)\notag\\
  & -\frac{1}{4}\delta_{\mu\nu}G_{\sigma\rho}^a(x, \tau)G_{\sigma\rho}^a(x, \tau)\\
  E(x,\tau)&=\frac{1}{4}G_{\sigma\rho}^a(x, \tau)G_{\sigma\rho}^a(x, \tau) .
\end{align}
 We can write
\begin{equation}
E_{ss}(x,\tau)=\frac{1}{4}\left (U_{ii}(x,\tau)-U_{00}(x,\tau)\right )-2 E(x,\tau) .
\end{equation}
The flow dependence then reads, using the expansions of \cite{Suzuki:2013gza},
\begin{align}
E_{ss}(x,\tau)=\frac{c_T(\tau)}{4}&\left ( T^R_{ii}-T^R_{00}\right )\notag\\
& -\frac{c_E(\tau)}{2}\left \{ G_{\mu\nu}G_{\mu\nu}\right\}^R - c_{\mathbf{1}}(\tau),
\label{eq:Ess_tau}
\end{align}
with $T_{\mu\nu}^R$ the renormalised field strength tensor and $\left \{ G_{\mu\nu}G_{\mu\nu}\right\}^R$ the renormalised version of $ G_{\mu\nu}G_{\mu\nu}$. The coefficients can be expanded perturbatively as
\begin{align}
    c_T(\tau)&=g_0^2+O\left(g^{\overline{\rm{MS}}}\left (\left(8\tau\right)^{-1/2}\right)\right) \\ c_E(\tau)&=1+O\left(g^{\overline{\rm{MS}}}\left (\left(8\tau\right)^{-1/2}\right)\right),
\end{align}
with $g_0$ the bare coupling and $g^{\overline{\rm{MS}}}$ the running coupling in the $\overline{\rm{MS}}$ scheme (see also \cite{DelDebbio:2013zaa}). The coefficient $c_{\mathbf{1}}$ is a mixing with unity and is set to $c_\mathbf{1}(\tau)=\langle E(\tau,x^{bulk})\rangle$ where by $x^{bulk}$ we mean the value in the centre of the lattice in the case of OBC. This sets the vacuum expectation of the trace of the energy-momentum tensor to zero \cite{Husung:2017qjz}. Equation \eqref{eq:Ess_tau} allows to obtain a renormalised quantity to study the screening mass,
\begin{equation}
\Delta E_{ss}^r =\lim_{\tau\to 0} \left( E_{ss}^r(\tau)-E_{ss}^{r_{bulk}}(\tau)\right ).
\end{equation}

\begin{figure}
\centering
  \includegraphics{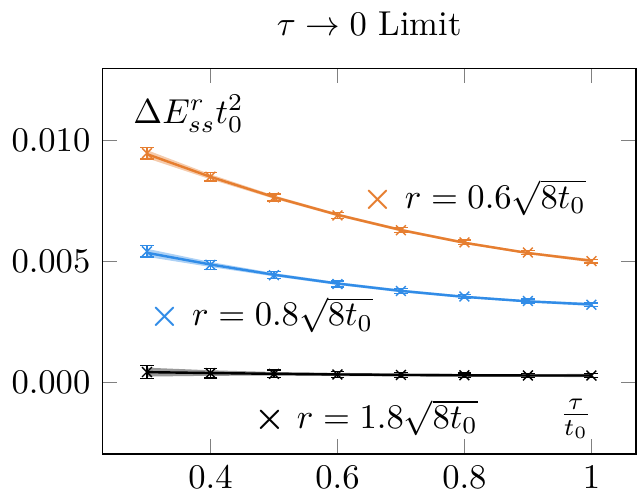}
    \caption{ Zero flow-time extrapolation of the continuum extrapolated $(E^{r}_{ss}-E^{bulk}_{ss})$.}
  \label{fig:lim_corrs_flowtime}
\end{figure}

\begin{figure}
\centering
  \includegraphics{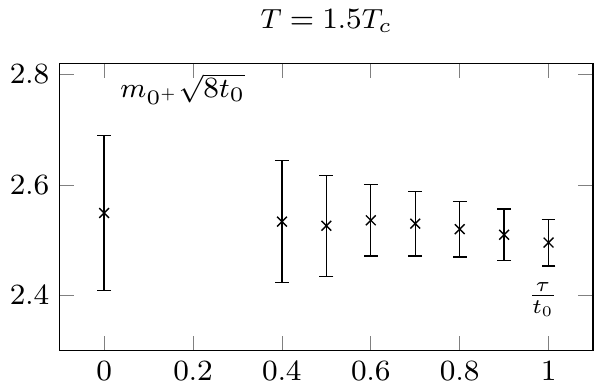}
  \caption{ Measured screening masses at different flow-times, at $T=1.5T_c$. They agree within statistical uncertainties. We clearly see the noise reduction associated with the Wilson flow.   The relatively small value of the scalar screening mass allows for a precise measurement and a zero flow-time extrapolation. }
  \label{fig:screening_masses}
\end{figure}

The zero flow-time extrapolation is shown in the top \modif{left} plot of figure \ref{fig:screening_proc}. To perform the extrapolation, we used a quadratic fit and checked that the result was insensitive to higher order corrections. An example at fixed radii is shown on figure \ref{fig:lim_corrs_flowtime}. As expected, the extracted screening mass is compatible with the one obtained at other flow-times, as is shown in figure \ref{fig:screening_masses}.

\begin{figure}
\centering
  \includegraphics{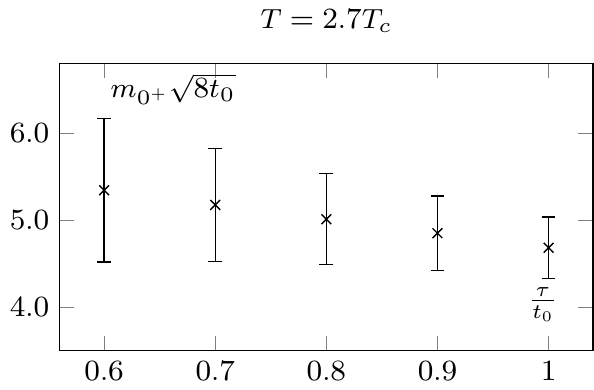}
  \caption{Screening mass at different flow times at $T=2.7T_c$. A larger mass is associated with larger uncertainties, see figure \ref{fig:screening_masses}.}
  \label{fig:screening_masses_2_7_Tc}
\end{figure}

We also extracted the screening mass at $T=2.7T_c$, but did not extrapolate to zero flow-time; this is shown in figure \ref{fig:screening_masses_2_7_Tc}. Note that the mass is noticeably larger at $2.7T_c$ than at $1.5T_c$; see section \ref{sec:discuss} for a discussion. Consistently, the errors are also larger at $2.7T_c$. It also explains why we did not proceed to a zero flow-time extrapolation. As we may see, the signal quickly worsens at small flow-time and the noise reduction associated to the flow is crucial to extract the mass. It is thus extracted at $t_0$.

\subsection{Pseudo-scalar Screening Mass}
\label{sec:pseudoscalar_mass}
Upon considering different operators, this method allows us to access the mass of the screening states of different quantum numbers. In this section, we will proceed with the mass determination of the pseudo-scalar screening state. One of the first continuum operators which come to mind and couples to the pseudo-scalar sector is the topological charge density

\begin{align}
  q(x)=\frac{1}{32\pi^2}\epsilon^{\mu\nu\rho\sigma}\Tr{G_{\mu\nu}G_{\rho\sigma}}
\end{align}

\begin{figure*}
\centering
  \includegraphics{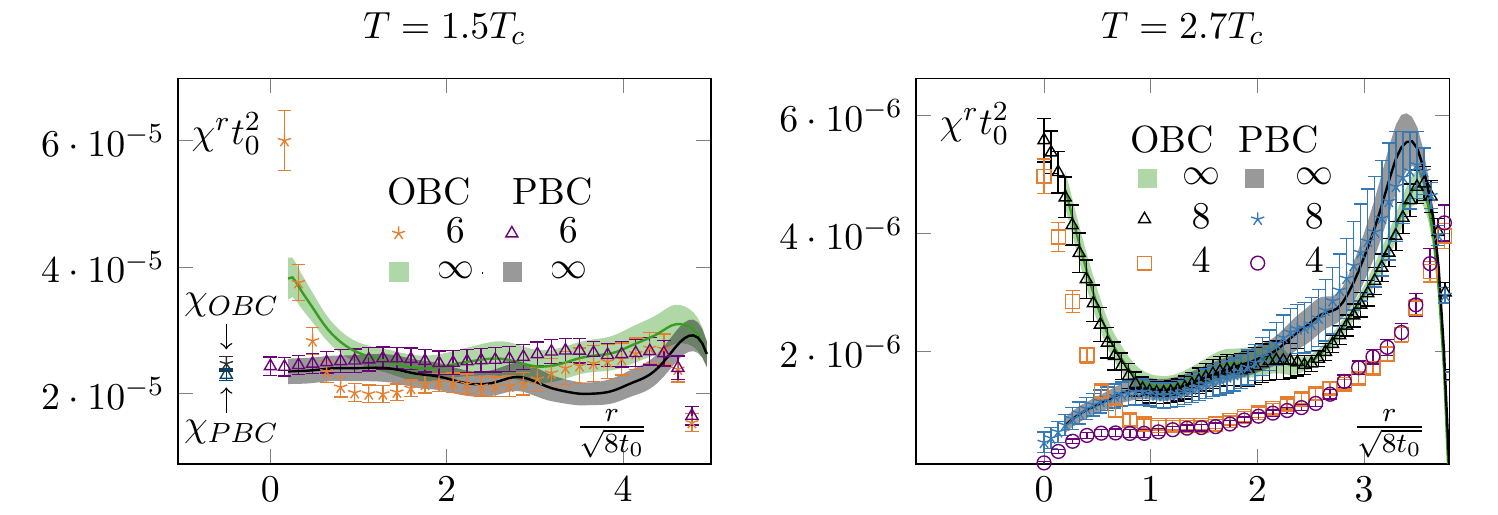}
  \caption{ \modif{Topological charge density square.}. The legend's labels correspond to $N_t$. For readability, we show only a subset of our data. At $T=1.5T_c$ (\textbf{left-hand side}), everything behaves as expected.  \modif{The topological charge density square} converges when integrated from the bulk and saturates to a constant value, which we can identify with the topological susceptibility. The OBC have the same bulk behaviour but suffer from exponentially suppressed contributions from the boundary states. The $T=2.7T_c$ case (\textbf{right-hand side}) is more interesting. We see that even the PBC charge density does not saturate. It can be interpreted as an indication of topological freezing, as it is known that the charge density over a sub-volume is less autocorrelated than the total charge \cite{Schaefer2011}. The OBC presents a similar pattern, calling for a more careful analysis of their autocorrelation time. }
  \label{fig:pseudo-scalar}
\end{figure*}

Unfortunately, we cannot proceed with its integrated average, as we did in section \ref{sec:scalar_mass} with the energy density, as $\langle Q\rangle=0$ in our case, with $Q$ the topological charge (eq. \eqref{eq:topCharge}); in other words we are in the sector $\theta=0$, with $\theta$ the QCD $\theta$-angle. To circumvent this issue, we consider the two-point function of $q$ over different sub-volumes,
\begin{align}
  \chi^r&\equiv\langle q^{2}\rangle^r = \frac{1}{N_yN_zN_t}\frac{1}{ \left (N_x-2r\right )}\langle\overline{q}^2(r)\rangle ,\label{eq:av_r}
\end{align}
\modif{where we defined an averaged $\overline{q}$, \begin{equation}
  \overline{q}(r)=\sum_{x=r}^{N_x-r-1}\sum_{y=0}^{N_y-1}\sum_{z=0}^{N_z-1}\sum_{t=0}^{N_t-1} q(x,y,z,t).
\end{equation}
In other words, $\chi^r$ is the average of topological charge square over a sub-volume.}  As the notation suggests, this quantity is related to the topological susceptibility, see section \ref{sec:top_susc}. \footnote{\modif{Note however that it is not the same quantity as is often used to study the topological susceptibility with OBC. Often, one considers the point-to-all two-point function of the topological charge, with the source far from the boundary (one can even average over different sources), see \cite{Bruno:2014ova} for more details.}} We show the $r$ dependence of this quantity in figure \ref{fig:pseudo-scalar} for various ensembles (some ensembles were omitted for the sake of clarity). Let us start discussing the ones at $1.5T_c$ (left-hand side of figure \ref{fig:pseudo-scalar}). As expected, we see again a boundary zone in the case of OBC and a saturation away from it. In the very centre of the lattice, $\chi^r$ displays a characteristic \say{bump}. This feature is inherited from the behaviour of the correlator $\langle q(x)q(0)\rangle$ around $x=0$ (see \cite{Giusti:2018cmp} for a detailed discussion).

The results at $2.7T_c$ display the same global features as the ones at $T=1.5T_c$, with a notable exception: $\chi^r$ does not completely saturate; we observe a drift in its plateau value. We understand this effect as a manifestation of topological freezing (the lattices at $2.7T_c$ are finer than the one at $1.5T_c$), see section \ref{sec:top_susc} for a discussion.

\begin{figure}
\centering
  \includegraphics{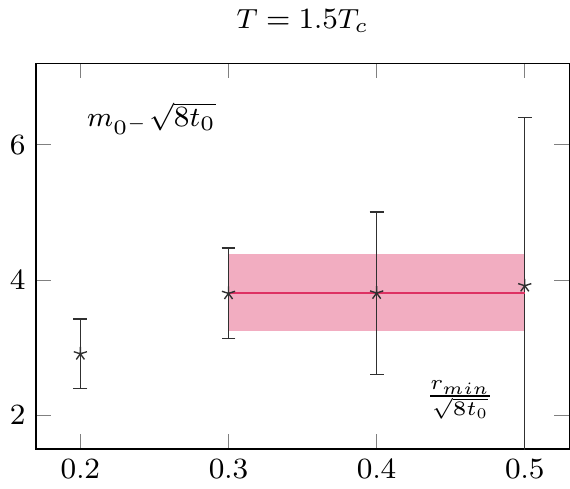}
  \caption{Extraction of the pseudo-scalar screening mass from the boundary pollution at $T=1.5T_c$. The $x$-axis is the radius at which we start our single-exponential fit. We extract the mass from the plateau value.  }
  \label{fig:pseudo-scalar_mass}
\end{figure}

\begin{figure}
\centering
  \includegraphics{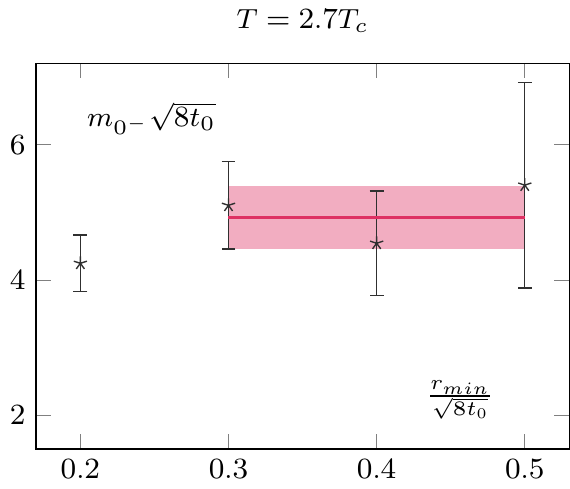}
  \caption{Extraction of the pseudo-scalar screening mass from the boundary pollution at $T=2.7T_c$. We observe a milder temperature dependence than in the scalar sector, see figure \ref{fig:pseudo-scalar_mass}. }
  \label{fig:pseudoscalar_mass_2_7_Tc}
\end{figure}

In all cases, to extract the screening masses, we are only interested in the exponential decay from the boundary. We use the same strategy as in the previous section. As the pseudo-scalar is heavier, we perform the extraction at flow-time  $t_0$ to have a good signal to noise ratio; as in the case of the scalar mass at $2.7T_c$, the signal quickly deteriorates for smaller flow-times. We show the results in figure \ref{fig:pseudo-scalar_mass} and \ref{fig:pseudoscalar_mass_2_7_Tc}. The errors are comparable to the ones obtained for the scalar at $2.7T_c$, as the masses are of similar magnitude. We also checked that the masses are (within the statistical uncertainties) independent of the maximal radius used for the fit, as long as this radius is taken within the plateau region of $\chi^r$.

\subsection{Discussion}
\label{sec:discuss}
\begin{figure*}
\centering
  \includegraphics{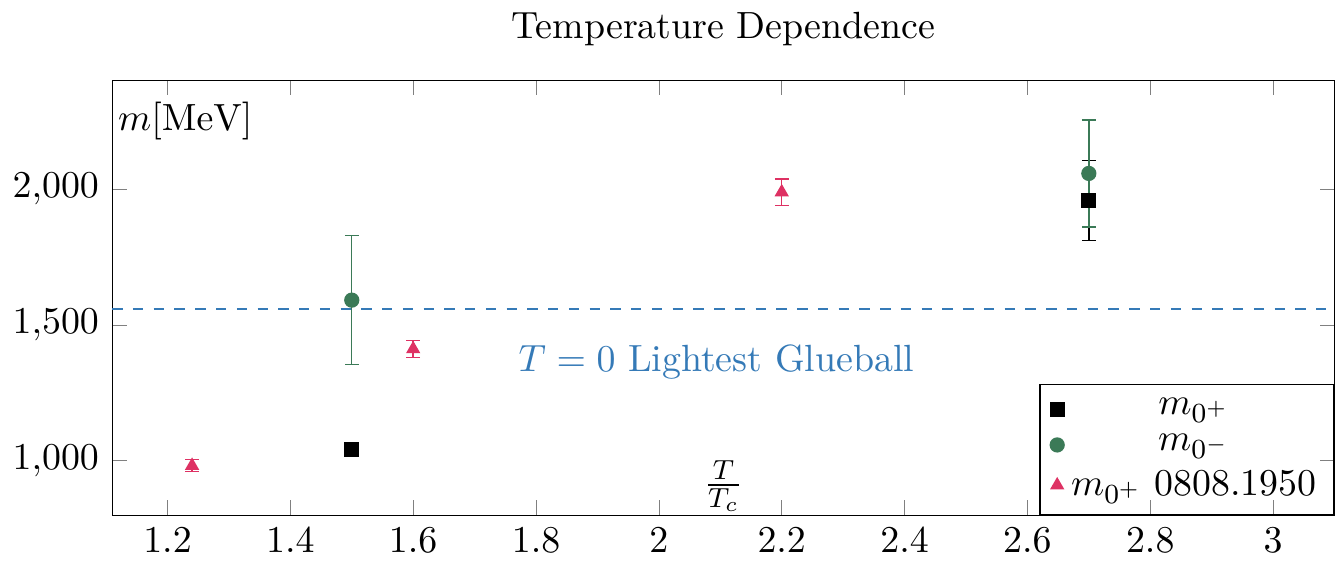}
  \caption{Summary plot for the screening masses  in [MeV]. Only statistical errors are shown. Firstly, as expected, the scalar screening mass is the lightest of the two states. Then, the behaviour of the masses is consistent with the behaviour of the boundary zone. At $T=1.5T_c$, the scalar screening mass is lighter than the $T=0$ lightest glueball. At $T=2.7T_c$ it is heavier. The behaviour of the pseudo-scalar mass is also consistent; from a large mass gap between the two channels at $T=1.5T_c$, we move to an almost degeneracy at $T=2.7T_c$, which is a signal of  dimensional reduction. On this figure, we also show the fixed lattice spacing results of \cite{Meyer:2008dt}. The $15\%$ discrepancy can most likely be attributed to systematic uncertainties (fixed lattice spacings, finite volume effects and conversion to physical units), even  though a systematic difference between our methods cannot be excluded.}
  \label{fig:temp_dep}
\end{figure*}

All  masses determined in this study are shown in physical units in figure \ref{fig:temp_dep}. As expected, being less symmetric, the pseudo-scalar state is heavier than the scalar state. Whilst certainly present at $T=1.5T_c$,  the difference is not statistically significant at $2.7T_c$. This is an indication of dimensional reduction; at high temperature, the scalar and pseudo-scalar are expected to become degenerate \cite{Datta:1998eb}.

On the same plot, we also show the values obtained in  \cite{Meyer:2008dt} by measuring the asymptotic behaviour of the energy density. The qualitative behaviour is the same but we observe a  shift of about $15\%$. Even if part of this discrepancy can presumably be explained by the fact that the results of \cite{Meyer:2008dt} are at fixed lattice spacings and other systematics, an intrinsic difference between the two methods cannot be excluded.

Setting this aside, the data of \cite{Meyer:2008dt} indicates that the main contribution to the pseudo-scalar mass is linear in $T$, as would be expected from  perturbation theory at high temperatures. Taking this for granted, we can estimate that the scalar screening mass becomes heavier than the lightest glueball at around $2T_c$. This should correspond to the temperature at which the boundary zone becomes strictly smaller than the zero temperature one. And indeed, the fact that the scalar screening mass at $1.5T_c$ is lighter than the lightest $T=0$ glueballs and that the $T=2.7T_c$ scalar screening mass is heavier is  consistent with what was reported in figure \ref{fig:bczone} about the length of the boundary zone.

\section{Topological Susceptibility}
\label{sec:top_susc}

\begin{figure}
  \centering
  \includegraphics{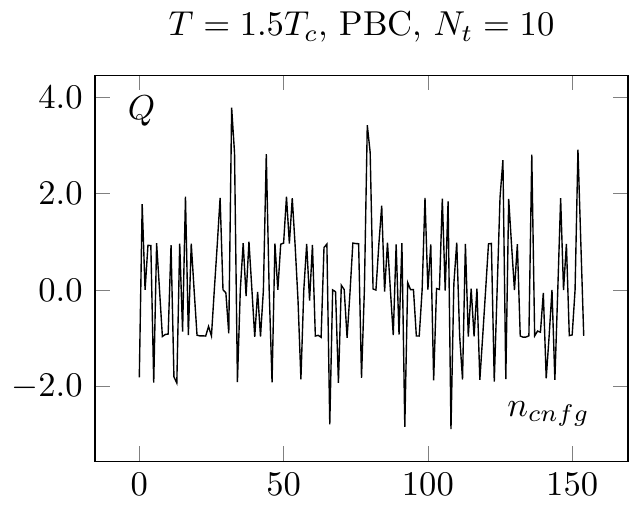}
  \caption{\textbf{Upper plots:} Topological charge history for PBC at $1.5 T_c$. Every configuration is separated by $500$ sweeps (see section \ref{sec:OBC}). The different topological sectors are well sampled.}
  \label{fig:freezing}
\end{figure}

\begin{figure}
  \centering
  \includegraphics{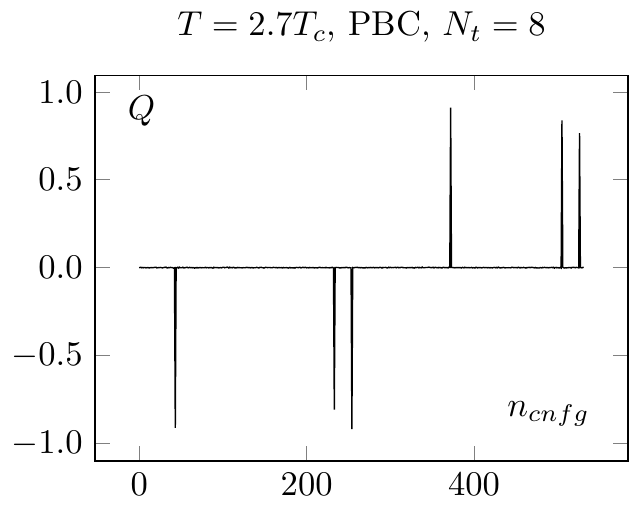}
  \caption{Topological charge history for PBC at $2.7 T_c$. We observe clear signs of topological freezing at $T=2.7T_c$. A very rough estimation gives $\tau_{auto}>300$. This confirms the behaviour observed in figure \ref{fig:pseudo-scalar}.}
  \label{fig:freezing_2_7_Tc}
\end{figure}

Actually, the \modif{topological charge density square} presented in section \ref{sec:pseudoscalar_mass} also allows us to extract a value for the topological susceptibility. Indeed, as the continuum topological susceptibility is defined to be

\begin{align}
  \chi&={\dd^2 E(\theta)\over \dd\theta^2 }\\ \label{def:chi}
  &=\frac{1}{V}\frac{1}{Z}\int \mathcal{D} A \left( \int \dd^4 x \int \dd^4 y \ q(x)q(y)\right ) e^{-S[A]},
\end{align}
with the $E(\theta)$ the vacuum energy at non-zero $\theta$,\footnote{We recall that Euclidean $SU(3)$ gauge theory at non-zero $\theta$ can be described by the Lagrangian $\mathcal{L}=L^{SU(3)}_{\theta=0}+i\theta q$.}
\begin{equation}
  E(\theta)=-\frac{1}{V}\ln\mathcal{Z},
\end{equation}
we expect the plateau value of $\chi^r$ to give the topological susceptibility.

 We show the obtained values in figure \ref{fig:pseudo-scalar}. The  $1.5T_c$ case is the most straightforward and leads to a clean signal. \modif{We perform a global fit on our three ensembles of the type $f(a) = c_1\cdot a^2 + \chi$ to remove the discretisation effect and extract the constant $\chi$. For PBC, we fit from the boundary up to $r_{max}=2\sqrt{8t_0}$. In the case of OBC, we excluded the data in the boundary zone. Correspondingly, we used values in the range $[1.3\sqrt{8t_0},2\sqrt{8t_0}]$.} It gives us measurements for the topological susceptibility
\begin{align}
    \chi_{OBC}(1.5 T_c)t_0^2&=2.47(15)\cdot10^{-5} \\
    \chi_{PBC}(1.5 T_c)t_0^2&=2.298(89)\cdot10^{-5},
\end{align}
which are in good agreement with $\chi(1.5 T_c) t_0^2=2.25(12)\cdot10^{-5}$ of \cite{Giusti:2018cmp} and  $\chi_t(1.5 T_c) t_0^2 \in [1.5\cdot10^{-5},4.4\cdot10^{-5}]$, the global fit of reference  \cite{Borsanyi:2015cka}. They are also consistent with the fixed lattice spacing results of \cite{Berkowitz:2015aua}.

\begin{figure}
  \centering
  \includegraphics{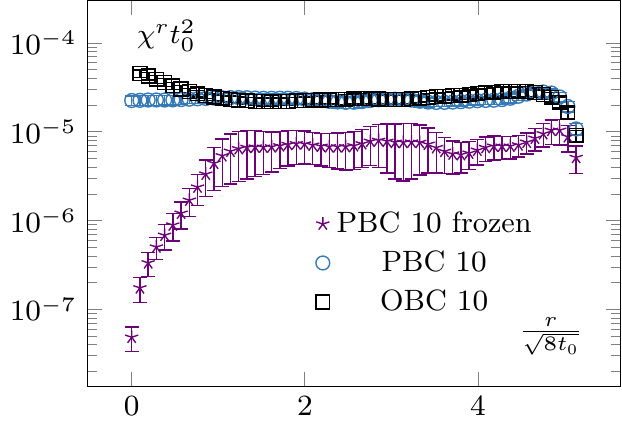}
  \caption{ \modif{Topological charge density square} on our finest lattice for OBC, PBC and $Q=0$ PBC configurations at $T=1.5T_c$. The numbers in the legend corresponds to $N_t$.}
  \label{fig:freezing_corr_1_5_Tc}
\end{figure}

At $2.7T_c$ the situation is more intricate, as it is not clear that $\chi^r$ saturates to a plateau. The trouble comes from two reasons. First, as it was already discussed in \cite{Bonati:2017woi} in the context of a toy model with OBC, $\chi^r$ can present a slow convergence as a function of $r$. This is what we see in the centre of the lattice. Then, the behaviour of the PBC configurations seems to indicate that our ensembles are partially frozen.

\begin{figure}
  \centering
  \includegraphics{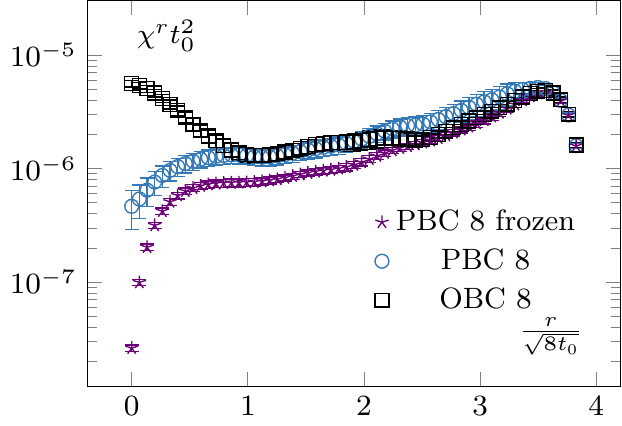}
  \caption{
 \modif{Topological charge density square} on our finest lattice for OBC, PBC and $Q=0$ PBC configurations at $T=2.7T_c$. We see that the correlator behaves similarly to the $Q=0$ restricted one at $T=1.5T_c$, see figure \ref{fig:freezing_corr_1_5_Tc}. The numbers in the legend corresponds to $N_t$.}
  \label{fig:freezing_corr_2_7_Tc}
\end{figure}

First, in figures \ref{fig:freezing} and \ref{fig:freezing_2_7_Tc}, we show the history of the topological charge for PBC. We observe a clear difference between the two temperatures. At $2.7T_c$, the topological transitions are highly correlated. Then, in figures \ref{fig:freezing_corr_1_5_Tc} and \ref{fig:freezing_corr_2_7_Tc},
we focus on our finest configurations at those two temperatures. We also display the results obtained when restricting ourselves to the $Q=0$ sector, i.e. by artificially freezing our lattices. Of course, at $1.5T_c$, the effect of freezing is drastic, as a lot of topological transitions are still to be expected. What is more interesting is the qualitative behaviour of $\chi_r$. For small sub-volumes, the value for the topological susceptibility is not so far from the unfrozen value but decreases for larger sub-volumes. It is consistent with the observation reported in \cite{Schaefer2011}, where it was observed that the topological charge measured on sub-volumes is less autocorrelated than the total charge. It is also intimately tied to the fact that the freezing of the topological charge is only a finite volume effect, one of the key ideas behind master field simulations \cite{Giusti:2018cmp,Luscher:2017cjh}, where very large volumes are generated and the ensemble average is obtained by summing over decorrelated sub-volumes. The fact that we do not get the correct value for the topological susceptibility is only due to our volumes being too small to perform sub-volume averages in fixed sectors.

This discussion can also be applied to the $2.7T_c$ case. However, there, the same kind of behaviour is present in the \say{unfrozen} case, which seems to indicate some partial freezing of our ensembles. This seems to be confirmed by the behaviour of the topological charge history of the PBC, which displays long correlations between jumps in topological sectors of the same sign; it shows correlations of at least $300$ configurations. Unfortunately, the OBC shows a similar kind of behaviour. This stresses the point that OBC are not a remedy to the topological freezing but only a potential improvement.

This discussion shows that no reliable estimate of $\chi(2.7T_c)$ can be extracted from figure \ref{fig:pseudo-scalar} without further investigations of the autocorrelations.
\modif{In particular, it confirms that even with these relatively large volumes,  an extraction of the topological susceptibility from $Q^2$ cannot be done reliably without a much larger statistics. }

\begin{figure}
  \includegraphics{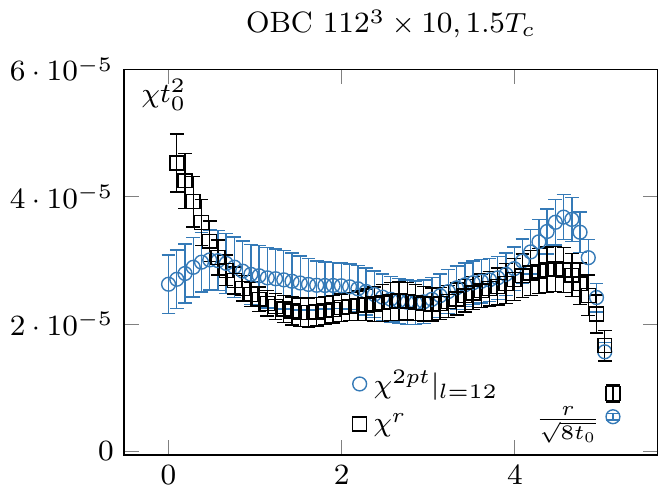}
  \caption{Topological charge density square $\chi^r$ versus topological charge integrated two-point function $\chi^{2pt}$. Both methods are consistent in their determination of the topological susceptibility. In this figure, $\chi^{2pt}$ was averaged over $24$ source points. }
  \label{fig:equiv_chi}
\end{figure}

Note also that the way we extracted the topological susceptibility in this work is not the only way to do it. One can also consider the point-to-all integrated two-point function of the topological charge, with the source far-away  from the boundary \cite{Bruno:2014ova}

\begin{align}
  \chi^{2pt}(r,l)=& \frac{1}{N_tN_yN_z}\frac{1}{2l}\label{eq:top_mattia}\\
  &\sum_{x_0=N_x/2-l}^{N_x/2+l-1}\sum_{x=r}^{N_x-r-1} q^{av}(x_0)q^{av}(x_0+x),\notag
\end{align}
with
\begin{equation}
  q^{av}(x)=\sum_{y=0}^{N_y-1}\sum_{z=0}^{N_z-1}\sum_{t=0}^{N_t-1} q(x,y,z,t).
\end{equation}
The first sum in \eqref{eq:top_mattia} is an average over sources that are far enough from the boundary while the second sum is a genuine integration. The quantity $2l$ is the number of source points which are averaged over. While we did not systematically study this quantity, we did check that our method is consistent with this definition. In figure \ref{fig:equiv_chi}, we show $\chi^r$ and $\chi^{2pt}(r,12)$ for our largest configurations at $T=1.5 T_c$. Note that the choice $l=12$ is presumably not optimal and the error bars associated to $\chi^{2pt}$ can presumably be reduced by tuning this parameter. We see that both methods are consistent; a careful study of their different systematics and how they relate is left as a potential interesting outlook.

\section{Conclusion}
\label{sec:conclu}

In this study, we started a first systematic investigation of OBC at high temperature. The main difficulty in dealing with OBC is the presence of boundary effects. In section \ref{sec:bczone}, we investigated the typical propagation length of these effects and compared it to the zero temperature results of \cite{Husung:2017qjz}. At $T=1.5T_c$, the boundary zone is larger than at $T=0$, while it is smaller at $T=2.7T_c$ and $T=3.0T_c$. These differences can be understood in terms of the temperature dependence of the mass of the lightest state in our system, namely the scalar screening mass. Actually, the boundary contamination gives us  means to measure this screening mass, giving results which are consistent with the already existing literature (see section \ref{sec:discuss}). In particular, we predict that the scalar starts to be heavier than the $T=0$ lightest glueball at around $T=2T_c$. It tells us that the use of OBC in the region $T\in [T_c,2T_c]$ is more delicate than at $T=0$ but becomes gradually easier at temperatures above $2T_c$. This is potentially useful as it is the interesting range of temperatures to measure the topological susceptibility, for example \cite{Moore:2017ond}. Moreover, we do not expect the situation to change drastically in full QCD, in the deconfined phase.

We also used the boundary effects in the pseudo-scalar channel to estimate the corresponding screening mass. We measured a sizable mass gap between the scalar and pseudo-scalar at $T=1.5T_c$. Moreover, we could  confirm  that this gap reduces at higher temperature, which is an expected signal of the dimensional reduction taking place at high enough temperatures.

As a by-product of the pseudo-scalar analysis, we could extract a precise measurement of the topological susceptibility at $T=1.5T_c$, which is in good agreement with the recent results of \cite{Giusti:2018cmp}. Finally, the same analysis at $T=2.7T_c$ exhibits some signs of topological freezing. A potential interesting outlook consists in studying quantitatively how the autocorrelation time depends on the lattice spacing and temperature and how it compares to the master field approach \cite{Giusti:2018cmp}.
\modif{Even so, it shows again that even with rather large volumes, the determination of the topological susceptibility is delicate.}
 This supports the recent efforts \cite{Giusti:2018cmp,Jahn:2018dke}, which have been undertaken to reassess the robustness of high-temperature studies of the topological susceptibility. In particular, a careful reconsideration of the finite size effects on its determination, even in the quenched case, is called for.

\section*{Acknowledgements}
The authors wish to thank M. Lombardo for an interesting discussion, A. Francis, M. Bruno and S. Sharma for comments on a previous version of this manuscript and particularly R. Sommer for his help and suggestions. A.F. thanks A. Rago, M, Hansen, F. Knechtli and  N. Husung for insightful discussions and A. Jaquier for his technical support. The work of A.F. is supported by the Swiss National Science Foundation. The authors acknowledge support by the Deutsche Forschungsgemeinschaft
(DFG, German Research Foundation) through the CRC-TR 211
\say{Strong-interaction matter under extreme conditions}– project number
315477589 – TRR 211.
Numerical calculations have been made
possible through PRACE grants at CSCS, Switzerland,
as well as grants at the GCS and at NIC-J\"ulich, Germany.
These grants provided access to resources on Piz Daint at CSCS and
JUWELS at NIC. Some of the numerical simulations have been performed on Fidis and Deneb, which are part of the EPFL HPC centre (SCITAS).

\FloatBarrier


\bibliographystyle{spphys}
\bibliography{biblio_OBC}
\end{document}